\documentclass[twocolumn]{aastex631}
\usepackage[flushleft]{threeparttable}
\usepackage{lmodern}
\usepackage{rotating}

\usepackage{afterpage}

\begin{document}

\title{Multi-Epoch Stability in the Rotational Modulation of the Planetary-Mass Companion Ross 458C}

\author[0000-0003-0192-6887]{Elena Manjavacas}
\affiliation{AURA for the European Space Agency (ESA), ESA Office, Space Telescope Science Institute, 3700 San Martin Drive, Baltimore, MD, 21218 USA}
\affiliation{Department of Physics and Astronomy, Johns Hopkins University, Baltimore, MD 21218, USA}
\correspondingauthor{Elena Manjavacas}
\email{emanjavacas@stsci.edu}
\author[0009-0001-7661-3108]{Jared Bull}
\affiliation{Department of Physics and Astronomy, Johns Hopkins University, Baltimore, MD 21218, USA}
\author[0000-0003-2446-8882]{Paulo Miles-Pa\'ez}
\affiliation{Centro de Astrobiolog\'ia, CSIC-INTA, Camino Bajo del Castillo s/n, 28692 Villanueva de la Ca\~nada, Madrid, Spain}
\author[0000-0001-5254-6740]{Natalia Oliveros-Gomez}
\affiliation{Department of Physics and Astronomy, Johns Hopkins University, Baltimore, MD 21218, USA}
\author[0000-0001-7356-6652]{Theodora Karalidi}
\affiliation{Department of Physics, University of Central Florida, 4111 Libra Dr, Orlando, FL 32816, USA}

\author[0000-0003-1487-6452]{Ben W. P. Lew}
\affiliation{Bay Area Environmental Research Institute, Moffett Field, CA 94035, USA}
\affiliation{NASA Ames Research Center, Moffett Field, CA 94035, USA}
\author[0000-0002-3726-4881]{Zhoujian Zhang}
\affiliation{Department of Physics \& Astronomy, University of Rochester, Rochester, NY14627, USA}
\author[0000-0003-2969-6040]{Yifan Zhou}
\affiliation{University of Virginia, 530 McCormick Road, Charlottesville, VA 22904, USA}
\author[0000-0003-3714-5855]{Daniel Apai}
\affiliation{Steward Observatory, The University of Arizona, 933 N. Cherry Avenue, Tucson, AZ 85721, USA}
\affiliation{Lunar and Planetary Laboratory, The University of Arizona, 1629 E. University Blvd., Tucson, AZ 85721, USA}

\author[0000-0003-2278-6932]{Xianyu Tan}
\affiliation{Tsung-dao lee institute and School of Physics and Astronomy, Shanghai Jiao Tong University, Shanghai 201210, China}



\begin{abstract}

Spectroscopic variability is detected in brown dwarfs across all spectral types, suggesting {atmospheric patchiness} in brown dwarf atmospheres. We present {JWST/NIRSpec time series observations} of the T8.0 planetary-mass companion Ross~458C. We compared {our derived} spectral variability and light curve {to those} obtained {using HST/WFC3 observations} more than 7.5~yr {before}  {and published by \cite{Manjavacas2019b}}. We {found} that the light curve of Ross~458C is remarkably similar in the two epochs in terms of shape and variability amplitude {potentially created by stable weather patterns}. {We measured a rotational period of {11.57$\pm$0.09~hr}, demonstrating that the previously reported HST/WFC3 period of 6.75$\pm$1.58 hr is most likely a {second-order} harmonic produced by an unresolved double-peaked rotational modulation, {likely due to the higher uncertainties of the HST/WFC3 light curve and the short time baseline of those observations.}} We measured the variability amplitude in the {white light curve created using only the  wavelength range covered by HST/WFC3 (1.10-1.63~$\mu$m), obtaining a value of 1.47$\pm$0.18\%},  which is consistent with that measured in the HST/WFC3 light curve. 

\end{abstract}




\section{Introduction}\label{sec:intro}

Most brown dwarfs and planetary-mass objects of all spectral types and ages show photometric or spectroscopic variability ({e.g. }\citealt{Radigan2012, Apai2013, Metchev2015, Yang2016, Biller2018, Vos2020}). One of the most likely causes of their variability is the existence of rotationally modulated heterogeneous clouds in their atmospheres, {rotating in and out of view} {\citep{Artigau2009, Apai2013}}. However, the handful of {L- and L/T transition brown dwarfs} monitored using the \textit{James Webb Space Telescope} (JWST, {\citealt{Gardner2023,Gardner2006,Rigby2023}}) with the NIRSpec {\citep{Bocker2023}} and the MIRI instrument {\citep{Wright2023}} have revealed that the cause of their variability might be due to a combination of factors in addition to or instead of heterogeneous clouds, like hot spots \citep{McCarthy2024} or thermochemical instabilities \citep{Chen2025, Oliveros-Gomez2026}.

 {Determining if the cause of the variability is the same for late-T brown dwarfs as for earlier spectral types is still an open question}. The silicate clouds present in mid-L and L/T transition brown dwarfs sink below the {observable photosphere} after the L/T transition \citep{Marley_Ackerman}. In fact, the spectra of most late-T dwarfs are well-fitted by cloudless spectra {(e.g. \citealt{Cushing2008})}. However, {there are} about a dozen of brown dwarfs with spectral types later than T5.0 that also show  variability \citep{Radigan2012, Buenzli2012, Metchev2015,  Manjavacas2019b, Miles_Paez2025}. In contrast to L/T transition brown dwarfs {that often show rapidly evolving light curves} {(e.g. 2MASS J13243553+6358281, 2MASS J01365662+0933473, and 2MASS J21392216+0220185 with spectral types of T2.0-T2.5, \citealt{Apai2017})}, {the variability for the few late-T dwarfs monitored until now is {often} sinusoidal, and very stable across epochs} \citep{Yang2016}, with faster rotational periods than earlier spectral type brown dwarfs. 2MASS~J22282889--431026 (2M2228--4310) is a clear example of this behavior \citep{Buenzli2012, Yang2016}.

For the case of Ross~458C, \cite{Burgasser2010_Ross} and \cite{Burningham2010} fitted its {IRTF/SpeX} near-infrared spectra {(1.0-2.5~$\mu$m)} to different atmospheric models, concluding that a model that includes silicate clouds reproduces the spectra of the object better than a cloudless model. 
In addition, \cite{Morley2012} fitted atmospheric models for late-T and Y-dwarfs that included sulfide clouds ($\mathrm{Na_{2}S}$), which reproduced {the SpeX 1.0-2.5~$\mu$m} near-infrared spectra of Ross~458C better than models including silicate clouds. {More recently, \cite{Zhang2021} attempted to fit cloud-free Sonora Bobcat models to Ross~458C's SpeX spectra, but found that these models produce too hot effective temperature values based on evolution models given this object's age and luminosity, further suggesting the presence of clouds. { \cite{Gaarn2023} performed a retrieval analysis in the same spectra also concluding the need of clouds}. {Finally, \cite{Meynardie2025} } fitted forward and retrieval atmospheric models to a JWST/NIRSpec spectrum (0.8-3.1~$\mu$m), also concluding the need to include silicate or sulfide clouds. }
These clouds might be one of the explanations for the spectroscopic variability found by \cite{Manjavacas2019b} using \textit{Hubble Space Telescope} (HST) and its \textit{Wide Field Camera 3} (WFC3) instrument {(PI: D. Apai, GO-14241)}.

In this letter we will focus on the comparison of the light curves obtained for Ross~458C with HST/WFC3 and the G141 grism published in \cite{Manjavacas2019b}, and the light curves obtained at similar wavelength ranges using JWST/NIRSpec (GO 5226, PI Manjavacas) to address its temporal evolution.

\section{Ross~458C}\label{sec:Ross458c}

Ross~458C \citep{Goldman2010} is a T8.0 substellar companion to the Ross~458AB binary system composed by a M0.5 and M7.0 star. Ross~458C has anomalous red colors in comparison to other substellar objects of the same spectral type ($J-K = -0.21\pm0.06$). The age of Ross~458C has been estimated by different methods using the binary host between 30-50~Myr \citep{Magazzu1993, Chabrier1996} and 400-800~Myr \citep{West2008}, although it is most likely slightly older than the Pleiades \citep{Burgasser2010} due to their deeper alkali lines. Using evolutionary models \citep{Chabrier, Baraffe} and an age estimation for the system, the estimated mass for Ross~458C is 12--25~$\mathrm{M_{Jup}}$, {placing it well within the cohort of low surface gravity companions}.



\section{Observations}\label{sec:observations}

Ross~458C was observed with the NIRSpec instrument \citep{Jakobsen2022} onboard JWST in June 7 and 8, 2025, as part of the Cycle 3 GO program 5226 (P.I. Manjavacas). We observed {our target} during 16.6~hr continuously in two separate non-interruptible visits of $\sim$8.3~hr each to avoid the data volume limit. Ross~458C has high proper motions {($\mu_{\alpha}$ = -628.7153~mas/yr, $\mu_{\delta}$ = -33.4718~mas/yr)}. To ensure that the target was centered in the S1600A1 slit, {we used pre-imaging using the EMIR instrument installed at the {GTC} Telescope in La Palma to obtain precise coordinates of the target few weeks before the NIRSpec observation}. In addition, to guarantee the success of the NIRSpec observation, we performed a Wide Aperture Target Acquisition (WATA) with the same slit. 
However, at the end of the first visit, the telescope returned to the initial pointing instead of staying at the corrected position after WATA. No WATA was performed before the second visit, leaving the spectral traces offset by $\sim$0.14" north of the slit center. {Thus, the spectra acquired during the second visit required a customized wavelength calibration (see Section \ref{sec:data_reduction})}.

We observed the target using the NIRSpec Bright Object Time Series (BOTS) template with the S1600A1 slit. We used the SUB512 array with the PRISM/CLEAR configuration, covering between 0.6-5.3~$\mu$m with a R$\sim$100. We applied no dithering to ensure the maximum stability of the observations. We used the NRSRAPID readout pattern, and 300 groups per integration, with a total of 440 integrations. In total we collected 880 spectra.

\section{Data Reduction}\label{sec:data_reduction}

The data were reduced using version 1.17.1 of the STScI JWST data reduction pipeline \citep{jwst_pipeline} and the {1322 CRDS version}. We ran Stage 1 using the default parameters on the \textit{*uncal.fits} files.
In Stage 2, we ran the default pipeline, running the NSClean algorithm to remove 1/f noise.  We used the \textit{*x1dints.fits} output files from Stage 2 for our analysis. 

For the spectra obtained in visit 2, we followed the same procedure as for the visit 1 spectra, but we modified the extraction window in Stage 2 to extract the offset spectra. For that, we adjusted the \texttt{jwst\_nirspec\_extract1d\_0006.json} reference file to move the extraction aperture in y-pixels measured in the \textit{*cal.fits} around the position of the trace, maintaining the same aperture width of 6 pixels. Since the visit 2 spectra were also slightly shifted in the dispersion direction, the wavelength calibration was not accurate, and it required manual wavelength calibration.  For that we cross-correlated the spectra obtained in the second visit using the first spectrum of Ross~458C from visit 1 with the 440 spectra obtained in visit 2 to find a wavelength shift that provides consistent wavelength calibration across all spectra. We used the \texttt{scipy.signal.correlate} Python function to look for correlations between the visit 2 spectra and the first spectrum of the series, and \texttt{scipy.signal.correlation\_lags} Python function to calculate the lags that best correlate the spectra. The calculated wavelength shift was one wavelength bin, corresponding to a shift between 0.005 and 0.020~$\mu$m depending on the wavelength, due to the spectrograph's wavelength-dependent spectral resolution.

\section{Results}\label{sec:resutls}


\subsection{Rotational Period Estimates}\label{sec:rotational_period}

We measured the variability of our target in the following wavelength ranges:  1.10-1.63~$\mu$m (white light curve {covering the same wavelength range as the HST/WFC3 observations}), 1.21-1.32~$\mu$m ($J$-band light curve), 1.54-1.60~$\mu$m ($H$-band light curve), and 1.35-1.43~$\mu$m (water-band light curve) as in \cite{Manjavacas2019b}. The light curves at {the rest of the wavelengths between 1.6--5.3~$\mu$m covered by the NIRSpec/PRISM will be presented in a future work} (Bull et al. 2026, in preparation). {Throughout the manuscript we will use this nomenclature to refer to the light curves at the specific wavelengths described above}. We binned the water-band every 8 points to increase the signal-to-noise. {Here we focus on the comparison between the HST/WFC3 and the JWST/NIRSpec light curves at the overlapping wavelengths.  }

In the following, we show the different methods used to measure the rotational period in all light curves {to cross-validate the resulting rotational period}: 

\begin{enumerate}

    \item \textit{{Truncated Fourier Series} plus systematic fit}: we fitted all the {median-normalized} light curves using the \texttt{curve\_fit} Python function inside \texttt{scipy.optimize} to a Fourier series of the form:

    \begin{equation}
    F(t) = \sum_{i=0}^{4} \left[ C_i \cos(2\pi f_i t) + S_i \sin(2\pi f_i t) \right]
    \end{equation}

    where $C_{i}$ and $S_{i}$ are the coefficients for each Fourier term, and $f_{i}$ are the frequencies ({1/$P_{i}$, where P is the rotational period}). We used the rotational period measured by \cite{Manjavacas2019b}, 6.75~hr, as the {initial guess} period for the Fourier fit. 
    
    In addition, we {include a function} that takes into account the systematics in the light curve introduced by the jitter on the pointing of the JWST telescope, reported before in other BOTS observations \citep{Rustamkulov2022, Alam2025}. To correct {for} these systematics we use the following {equation} as in the works mentioned:

    \begin{equation}
    {Sys (x,y,t)} = a_0 + a_1 x + a_2 y + a_3 t
    \end{equation}

    where $x$ and $y$ are the drift of the spectral trace in pixels with respect to the initial position, $t$ is time, and $a_{0}$, $a_{1}$, $a_{2}$ and $a_{3}$ are the fitted coefficients. We fit the light curve with both models: 

    \begin{equation}
    m(x, y, t) = {Sys(x, y, t)} \cdot F(t)
    \end{equation}

    {However, since we normalized the data previous to the fit ($a_{0} \sim 1.00$) and both the amplitude of the light curve and the magnitude of the systematics {are} very small ($a_{1}$, $a_{2}$ and $a_{3}$ $\sim$ 0.00), we can approximate the equation by}:

    \begin{equation} \label{eq:model}
    m(x, y, t) = {Sys(x, y, t)} + F(t)
    \end{equation}

    which is a linear equation much simpler to solve. By fitting simultaneously the Fourier function and a linear term, we remove a {linear} downtrend introduced most likely by systematics. In Fig. \ref{fig:light_curves} we show all the light curves after the correction, and the best Fourier function (Fig. \ref{fig:light_curves}, panels: a, b, e, f, {dashed black lines}), and the residuals (Fig. \ref{fig:light_curves}, panels: c, d, g, h, {gray residuals}). In Table \ref{table:combined_results} we show the best fit parameters to the systematic model and the Fourier fit. We used a Bayesian Information Criterion {(BIC, \citealt{KassRaftery1995})} to choose the model that optimizes the fit using the least number of free parameters. The BIC is defined as: $\text{BIC} = k \ln(N) - 2 \ln(\hat{L})$, where $k$ is the number of fitted parameters depending on the order of the Fourier function ({i = 0--4)}, $N$ is the number of data points (N = 880), and $2 \ln(\hat{L})$ represents the goodness of the fit. The lowest BIC provides the most optimal fitting model, which was in this case {for a Fourier function of i=1}. We obtain a period of 5.75$\pm$0.07~hr for the white light curve, 5.85$\pm$0.03~hr for the $J$-band light curve, 5.71$\pm$0.03~hr for the $H$-band light curve and 5.75$\pm$0.23~hr for the water band light curve, which are consistent with the rotational periods derived by \cite{Manjavacas2019b}. In Fig. \ref{fig:light_curves} we show the best fitting Fourier function to the corrected light curve from systematics {(dashed black line, panels a, b, e, and f)}. As observed in Fig. \ref{fig:light_curves}, {panels c, d, g, and h (gray residuals),} the residuals are not flat showing remaining structure, probably indicating that the Fourier fit is not optimal. In Table~\ref{table:combined_results} we show the standard deviation of the residuals for the fits to all light curves representing the goodness of each fit.

    \begin{figure*}[htbp]
    \centering
    \includegraphics[width=0.7\textwidth]{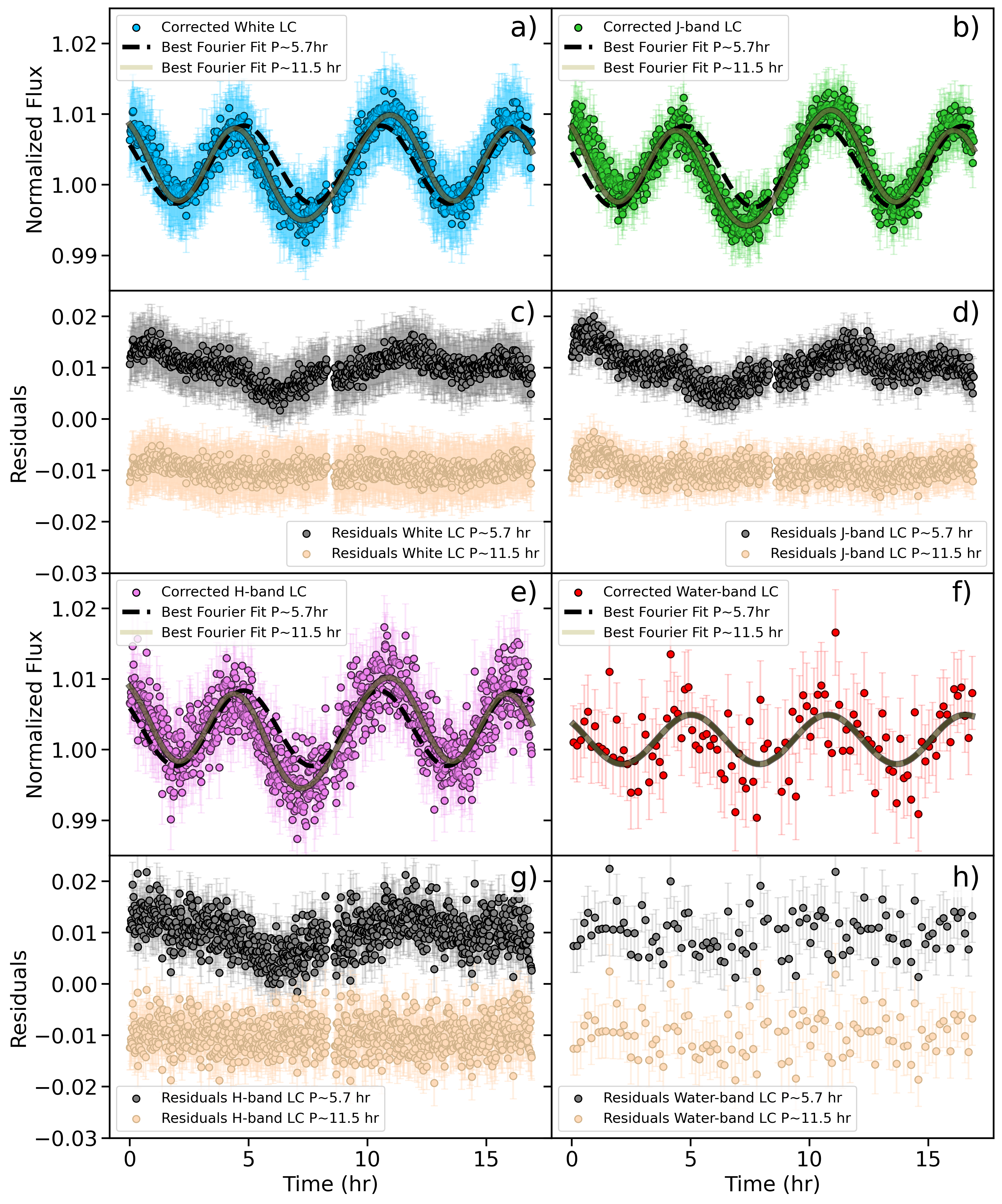}
    \caption{{White light curve (1.10-1.63~$\mu$m, blue, panel a)}, $J$- (green, panel b), $H$- (pink, panel e), and water-band (red, panel f) light curves, with their respective best fit models after being {corrected by} systematics (panels c, d, g, and h). The best Fourier fit with P$\sim$5.7~hr (dashed black line) and with P$\sim$11.5~hr (solid gray line). We also include in a panel below each light curve their respective residuals after subtracting their best fit model using a period {initial guess} of $P_{{initial guess}}$=5.7~hr (gray), and in light brown/peach color after subtracting their best fitting model with $P_{{initial guess}}$=11.5~hr. }
    \label{fig:light_curves}
    \end{figure*}

    \item \textit{Lomb-Scargle Periodogram}: We run a Lomb-Scargle periodogram \citep{Horne_Baliunas1986} to measure the rotational period on the corrected white, $J$-band, $H$-band and water corrected light curves. We calculate the uncertainties as explained in \cite{Horne_Baliunas1986}, and we use a {false alarm probability} of 0.001 \citep{Baluev2008} to discard the spurious peaks in the periodogram. The periodogram shows one prominent peak for all light curves at around between 5.7-5.8~hr for all the light curves. For the white light curve we obtain a period of 5.77$\pm$0.07~hr, for the $J$-band light curve a period of 5.86$\pm$0.10~hr, for the $H$-light curve a period of 5.70$\pm$0.07~hr, and for the water light curve a period of 5.77$\pm$0.28~hr, consistent with the results obtained using the Fourier fit.

    \item \textit{Bayesian Generalized Lomb-Scargle Periodogram}: We also run a Bayesian Generalized Lomb-Scargle Periodogram (BGLS) on all the corrected light curves. {BGLS was} found to be less biased when there are temporal gaps in data \citep{Mortier2015}, {and} uses the uncertainty in the data {to find the best period}. We calculated the uncertainties in the derived periods by fitting a Gaussian function to the mean peak and calculating the Full Width Half Maximum (FWHM). We obtained a rotational period of 5.77$\pm$0.03~hr for the white light curve, a period of 5.86$\pm$0.03~hr for the $J$-band light curve, a period of 5.70$\pm$0.03~hr for the $H$-band light curve and a period of 5.77$\pm$0.03~hr for the water light curve.


    \item \textit{Gaussian Processes (GP):} Finally we use a GP to measure the rotational period of the target. 
    To run GP in Python we use the function \textit{celerite2} \citep{celerite2}, using the \textit{RotationTerm} as the Covariance Matrix which is best suited for quasi-periodic signals. 
    We used a P=6.75~hr as a {initial period estimate} to compute the GP for all the light curves, {and a flat prior between 0.3 and 20~hr}. For the white light curve we obtained a bimodal distribution in the rotational period, with a peak at P=5.73$\pm$0.49~hr, and a similar peak at P=11.38$\pm$0.52~hr. For the $J$-band light curve we obtain also a bimodal distribution for the period with the higher peak at P=11.21$\pm$0.52~hr and a much smaller at P=5.90$\pm$0.56~hr. Similarly, for the $H$-band light curve we obtained a bimodal distribution for the period, being the highest peak at 11.38$\pm$0.44~hr, and a much smaller one at 5.83$\pm$0.63~hr. Finally, for the water band light curve we obtained a single peak distribution for the period with a value of P=5.77$\pm$0.46~hr.

\end{enumerate}

\begin{deluxetable*}{lcccccccccc}
\tabletypesize{\footnotesize}
\tablecaption{Best-fit Systematics and Fourier Model Parameters \label{table:combined_results}}
\tablecolumns{11}
\tablewidth{0pt}
\setlength{\tabcolsep}{3pt} 

\tablehead{
    \colhead{} & \multicolumn{10}{c}{Model Parameters}
}

\startdata
\cutinhead{Systematics Model Parameters ($a_n$)}
{Light Curve} & \multicolumn{2}{c}{$a_0$} & \multicolumn{2}{c}{$a_1$} & \multicolumn{2}{c}{$a_2$} & \multicolumn{4}{c}{$a_3$ ($10^{-4}$)} \\
\hline
\noalign{\smallskip}
White ($P_{5.7}$) & \multicolumn{2}{c}{1.000$\pm$0.001} & \multicolumn{2}{c}{-0.001$\pm$0.034} & \multicolumn{2}{c}{-0.001$\pm$0.105} & \multicolumn{4}{c}{$-3.89\pm1.00$} \\
$J$-band ($P_{5.7}$) & \multicolumn{2}{c}{1.003$\pm$0.001} & \multicolumn{2}{c}{-0.001$\pm$0.022} & \multicolumn{2}{c}{-0.001$\pm$0.069} & \multicolumn{4}{c}{$-3.59\pm0.65$} \\
$H$-band ($P_{5.7}$) & \multicolumn{2}{c}{1.003$\pm$0.001} & \multicolumn{2}{c}{-0.001$\pm$0.028} & \multicolumn{2}{c}{-0.001$\pm$0.087} & \multicolumn{4}{c}{$-3.78\pm0.83$} \\
$\mathrm{H_{2}O}$ ($P_{5.7}$) & \multicolumn{2}{c}{1.001$\pm$0.002} & \multicolumn{2}{c}{-0.001$\pm$0.113} & \multicolumn{2}{c}{-0.001$\pm$0.223} & \multicolumn{4}{c}{$-1.14\pm1.40$} \\
\hline
White ($P_{11.5}$) & \multicolumn{2}{c}{1.000$\pm$0.001} & \multicolumn{2}{c}{-0.001$\pm$0.034} & \multicolumn{2}{c}{-0.001$\pm$0.105} & \multicolumn{4}{c}{$-3.89\pm1.00$} \\
$J$-band ($P_{11.5}$) & \multicolumn{2}{c}{1.003$\pm$0.001} & \multicolumn{2}{c}{-0.001$\pm$0.022} & \multicolumn{2}{c}{-0.001$\pm$0.069} & \multicolumn{4}{c}{$-3.59\pm0.65$} \\
\cutinhead{Fourier Series Parameters}
{Light Curve} & {n} & {BIC} & {$f_{1}$} & {$C_{1}$} & {$S_{1}$} & {$C_{2}$} & {$S_{2}$} & {$C_{3}$} & {$S_{3}$} & {$std_{res}$} \\
 &  &  & ($10^{-2}$) & ($10^{-3}$) & ($10^{-3}$) & ($10^{-3}$) & ($10^{-3}$) & ($10^{-3}$) & ($10^{-3}$) & ($10^{-3}$) \\
\hline
\noalign{\smallskip}
White ($P_{5.7}$) & 1 & 467 & 17.3$\pm$0.2 & 2.27$\pm$0.46 & -4.76$\pm$0.34 & \dots & \dots & \dots & \dots & 3.583\\
$J$-band ($P_{5.7}$) & 1 & 3196 & 17.1$\pm$0.1 & 2.03$\pm$0.32 & -5.38$\pm$0.20 & \dots & \dots & \dots & \dots & 6.092\\
$H$-band ($P_{5.7}$) & 1 & 726 & 17.5$\pm$0.1 & 2.86$\pm$0.38 & -4.49$\pm$0.29 & \dots & \dots & \dots & \dots & 3.875 \\
$\mathrm{H_{2}O}$ ($P_{5.7}$) & 1 & 95 & 17.3$\pm$0.7 & 2.37$\pm$1.16 & -2.56$\pm$1.14 & \dots & \dots & \dots & \dots & 4.145 \\
\hline
\noalign{\smallskip}
White ($P_{11.5}$) & 3 & 369 & 8.64$\pm$0.07 & 2.47$\pm$0.30 & 1.09$\pm$0.29 & 2.55$\pm$0.46 & -5.32$\pm$0.33 & 1.01$\pm$0.27 & 0.54$\pm$0.28 & 2.971\\
$J$-band ($P_{11.5}$) & 3 & 2892 & 8.54$\pm$0.04 & 2.91$\pm$0.20 & 0.91$\pm$0.19 & 2.05$\pm$0.32 & -5.96$\pm$0.20 & 0.88$\pm$0.18 & 0.18$\pm$0.18 & 5.688 \\
$H$-band ($P_{11.5}$) & 3 & 510 & 8.70$\pm$0.06 & 2.97$\pm$0.25 & 1.23$\pm$0.25 & 2.42$\pm$0.39 & -5.35$\pm$0.27 & 0.92$\pm$0.23 & 0.66$\pm$0.24 & 3.112 \\
$\mathrm{H_{2}O}$ ($P_{11.5}$) & 1 & 95 & 17.3$\pm$0.7 & 2.37$\pm$1.16 & -2.56$\pm$1.14 & \dots & \dots & \dots & \dots & 4.146 \\
\enddata
\tablecomments{$std_{res}$ is the standard deviation of the residuals after subtracting the best fitting Fourier fit. {The white light curve is created using only the HST/WFC3 wavelength range (1.10-1.63 micron).}}
\end{deluxetable*}

\subsection{Determining the true rotational period}\label{section:true_period}

{Using different methods, we obtained two rotational periods that might be present in all light curves except the water-band light curve, one consistent with the rotational period reported by Manjavacas et al. (2019) ($P\sim5.7$ hr) and another of $P\sim11.5$ hr. The shorter period is recovered by the Fourier-series fits presented in Section 5.1.1 when optimized using \texttt{scipy.optimize.curve\_fit}, which can preferentially converge to a local solution associated with the initial period guess adopted. Broader exploration of the same Fourier-model parameter space reveals an alternative family of solutions near twice this period, consistent with the longer-period peak identified by the GP analysis (see Fig. \ref{fig:light_curves}, panels a, b, e, f, gray solid line). To determine which of the two periods is most likely its true rotational period, we use two procedures: 1) we fit again the Fourier functions as explained in the previous Section, having as the initial guess periods $P = 5.7$ hr and $P = 11.5$ hr. The period that provides the smallest residuals will most likely be its true rotational period, after using the BIC to choose the best fitting Fourier function. 2) We phase-folded each light curve using both periods to determine if the rotational profiles exhibited consistency across different cycles.}

In Fig. \ref{fig:light_curves} we show in panels c, d, g and h, the residuals of the best Fourier fits to our light curves using P=5.7~hr as {initial guess} (gray residuals), and P=11.5~hr as {initial guess} (peach/light brown residuals). The standard deviation of the gray residuals ($\mathrm{P_{{initial guess}}}$ = 5.7~hr) and the peach/light brown residuals ($\mathrm{P_{{initial guess}}}$ = 11.5~hr) are shown in Table~\ref{table:combined_results}. The standard deviations for the Fourier fit using $\mathrm{P_{{initial guess}}}$ = 11.5~hr as the {initial guess} period {are smaller, as well as the} BIC, suggesting that P$\sim$11.5~hr is the true rotational period. The exact rotational period derived using this method is: P = 11.57$\pm$0.09~hr.

\begin{figure*}[htbp]
\centering
\includegraphics[width=0.6\textwidth]{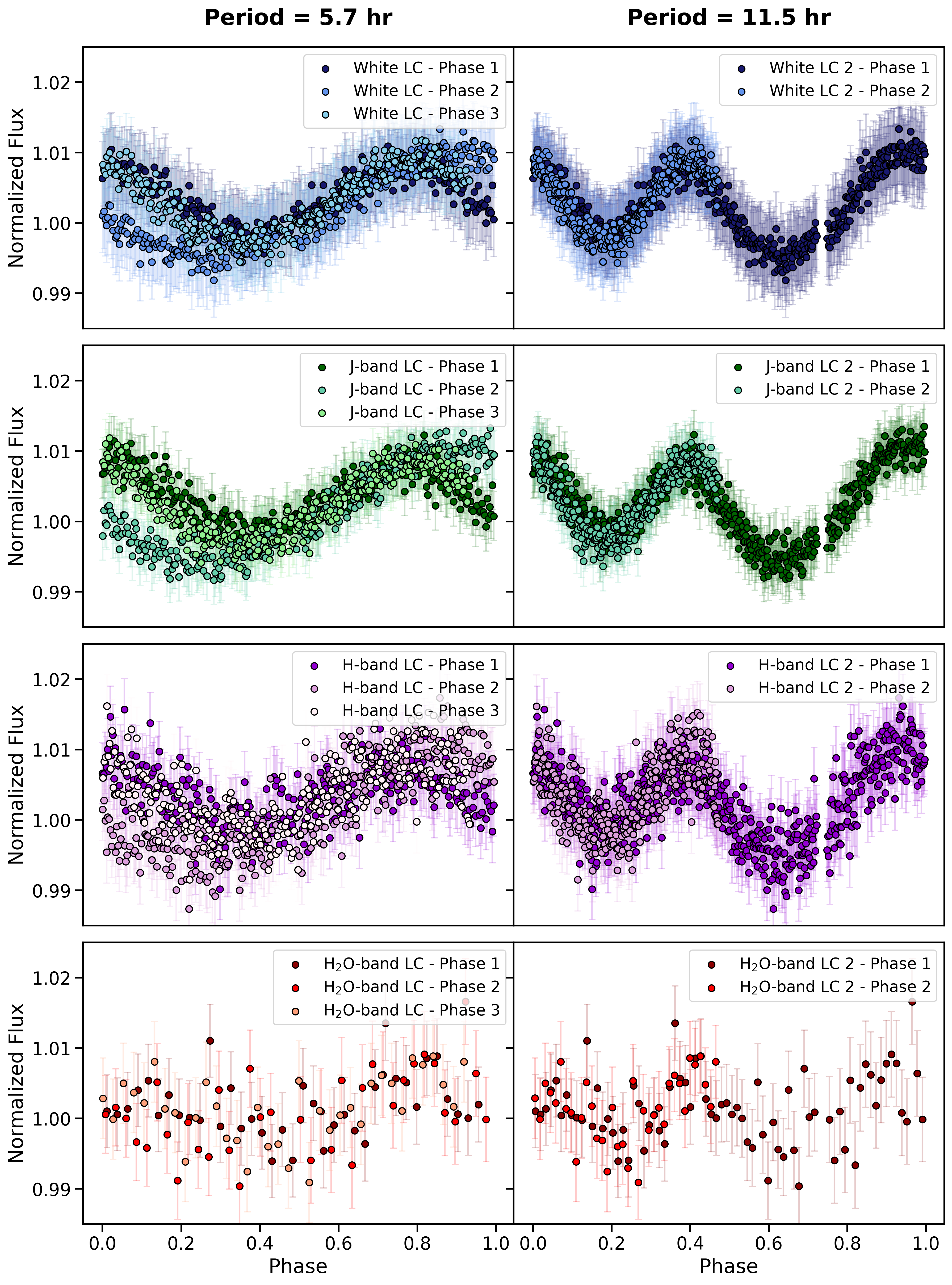}
\caption{Phase folded light curves using the P$\sim$5.7~hr rotational period (left panels) and the P$\sim$11.5~hr rotational period (right panels). {As a reminder, the white light curve is created using only the HST/WFC3 wavelength range (1.10-1.63 micron)}. We show in different shades of the same color the light curves corresponding to the different rotations, assuming both rotational periods.}
\label{fig:phase_folded_light_curves}
\end{figure*}

In Fig. \ref{fig:phase_folded_light_curves} we show all the light curves phase-folded using the P=5.7~hr period (left panels), and the P=11.5~hr period (right panels). As we can observe in Fig. \ref{fig:phase_folded_light_curves}, the light curves folded with the P=11.5~hr period overlap much better in all the rotations. {Thus, we adopt the period derived using the intial guess of P = 11.5~hr measured in the white light curve: {11.57$\pm$0.09~hr}}.

\subsection{Spectral Variability}\label{sec:spectral_variability}

\begin{figure}[htbp]
\centering
\includegraphics[width=0.48\textwidth]{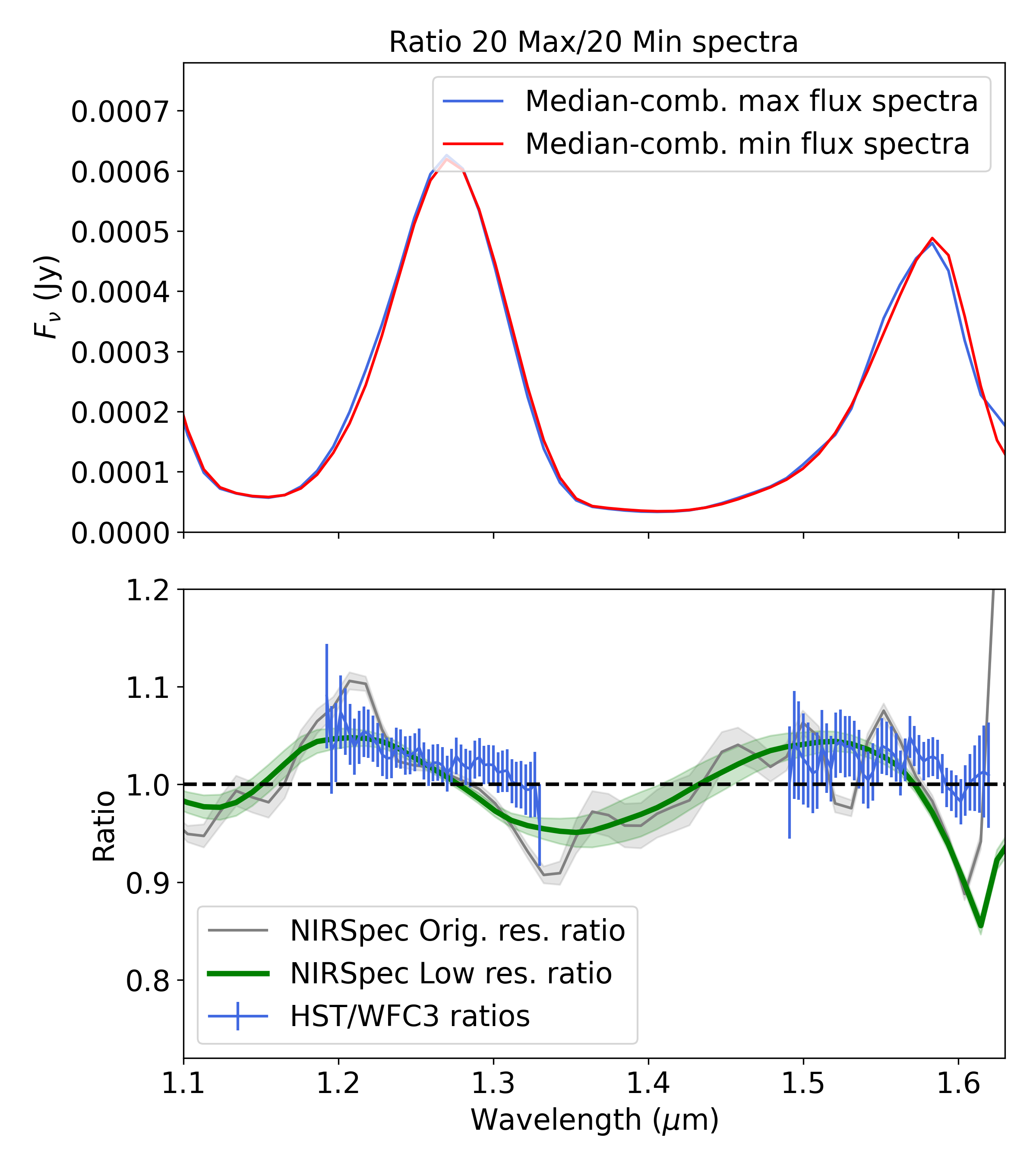}
\caption{Top panel: median combination of the {20} maximum flux (blue) and 20 minimum flux spectra (red). Bottom panel: spectral ratio of the JWST/NIRSpec spectra between 1.1 and 1.62~$\mu$m before smoothing (gray line), and after smoothing (green line). We also show the ratios obtained by \cite{Manjavacas2019b} using HST/WFC3 spectra (blue).}
\label{fig:spectral_variability}
\end{figure}

Following \cite{Manjavacas2019b} we measured the spectral variability via the ratio between the maximum flux spectrum of the series and the minimum flux spectrum, and we compared it to the results obtained by \cite{Manjavacas2019b} using HST/WFC3 spectra. We used as the "maximum flux spectrum" a median of the 20 spectra with the maximum flux in the light curve, obtained between 10.52~hr and 10.88~hr, and the "minimum flux spectrum" is a median of the 20 spectra with the minimum flux in the light curve, obtained between 6.85~hr and 7.21~hr. {In Figure \ref{fig:spectral_variability}, top panel, we show the {20} median combined maximum (blue) and {20} minimum spectra (red)}.  {In Fig. \ref{fig:spectral_variability}, bottom panel, we show the ratio between them (gray line), and the same after smoothing the resolution  using the \texttt{lowess} Python function inside the \texttt{statsmodels} package with a local fraction of frac=0.03, meaning each localized linear regression was computed using a neighborhood window constrained to exactly 3\% of the total sample size. The ratio between the median combined maximum flux spectra and minimum flux spectra shows the wavelength-dependent variability}. The peak-to-peak variability amplitudes measured for all light curves are: 1.47$\pm$0.19\% in the white light curve, 1.48$\pm$0.23\% in the $J$-band light curve, 1.60$\pm$0.49\% in the $H$-band light curve and 1.06$\pm$0.32\% in the water band light curve. 
{In addition, {in Fig. \ref{fig:spectral_variability}} we include the ratios published by \cite{Manjavacas2019b} with its uncertainties in blue, which are consistent with the ratios presented here.} 

\section{Discussion}\label{sec:discussion}

\subsection{{Comparison to HST/WFC3 light curve shapes, amplitudes and periods}}\label{sec:comparison_HST_var}

\begin{figure}[htbp]
\centering
\includegraphics[width=0.49\textwidth]{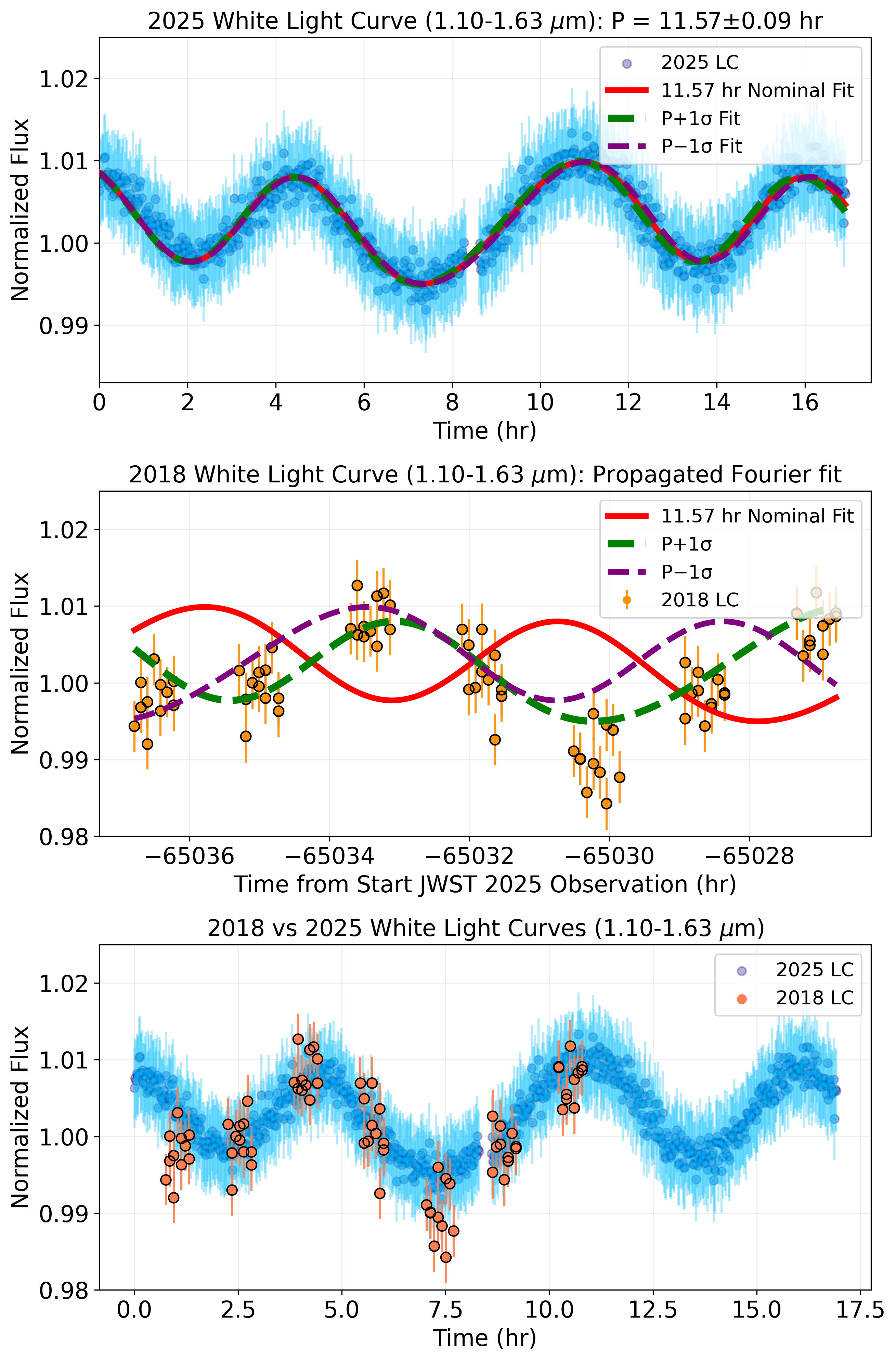}
\caption{{Top panel: {2025 White light curve covering the same wavelength range as the HST/WFC3 White light curve obtained in 2018} with its best Fourier function fit (P = 11.57$\pm$0.09~hr, red line). The same fits using the derived period plus its 1$\sigma$ uncertainty (green), and the derived period minus its 1$\sigma$ uncertainty (purple). Medium panel: {2018} HST/WFC3 White light curve with the best Fourier fits calculated using the JWST/NIRSpec White light curve {(1.10-1.63~$\mu$m)} propagated to the epoch of the HST/WFC3 light curve. {Bottom panel: Cross-correlated 2018 HST/WFC3 and 2025 JWST/NIRSpec 1.10-1.63~$\mu$m light curves overplotted}.}}
\label{fig:comparison_HST}
\end{figure}

\cite{Manjavacas2019b} obtained a rotational period of 6.75$\pm$1.58~hr which is {similar to} the rotational periods that we obtained with some of the methods explained in Section \ref{sec:rotational_period}. However, as explained in Section \ref{section:true_period}, this period is most likely a second harmonic of the true rotational period.
There are two reasons that might explain why this is the case: the first is the time coverage of that light curve. The HST/WFC3 light curve only spanned {10.11~hr}, which did not fully {covered} one true rotational period {(11.57$\pm$0.09~hr)}. The second reason is that Ross~458C is relatively faint for HST ($J$-band = 16.7~mag and $H$-band = 17.0~mag); {thus, the two different height peaks per rotational period that we observe in the NIRSpec light curve are not distinguishable in the HST/WFC3 light curve} (see Fig. \ref{fig:comparison_HST}, {bottom panel, orange light curve}). This is likely the same reason why we do not find the $\sim$11.5~hr rotational period in the water-band light curve in the JWST/NIRSpec data.

{Regarding the comparison between variability amplitudes in both epoch light curves}, if we measure the variability amplitude of the HST/WFC3 light curve using a median of the four points around the maximum and the minimum of the light curves, we obtain a variability amplitude of 1.61$\pm$0.49\% for the white-band light curve, 2.04$\pm$0.38\% for the $J$-band, and 2.30$\pm$0.51\%, which is consistent with the variability amplitudes we measure in the JWST/NIRSpec light curves presented in Section~\ref{sec:spectral_variability}. 

{To demonstrate the potential light curve stability, {in Fig. \ref{fig:comparison_HST} (top panel) we show the white-band JWST/NIRSpec light curve of Ross~458C obtained in June 2025, and in the middle panel we show the HST/WFC3 light curve obtained in January 2018. We overplot in both panels the best Fourier function fit as found in Section~\ref{sec:rotational_period} (solid, red line) using the derived period from the fit with a 11.5~hr initial guess (P = 11.57$\pm$0.09~hr). In addition, we overplot the same Fourier fit using the derived period plus its derived uncertainty (P = 11.66~hr, dashed purple line), and minus its derived uncertainty (P = 11.48~hr, dashed purple line)}. }

{As shown in Fig. \ref{fig:comparison_HST}, middle panel, the rotational period derived plus 1$\sigma$ (P = 11.66~hr) fits both JWST/NIRSpec and HST/WFC3 light curves simultaneously. Finally, in Fig. \ref{fig:comparison_HST}, bottom panel, we overplot both light curves after aligning them using cross-correlation. This plot further shows the similarity in amplitude and shape of both epoch light curves. We cannot exclude that the light curve might have evolved between the two epochs, but we can conclude that both light curves show similar amplitudes and periods.}

\subsection{Possible Causes of the Stable Long Term Variability} \label{sec:comparison_phase_folded}

In contrast to earlier spectral type brown dwarfs, which usually show light curves that evolve within a few rotational periods {\citep{Apai2017,Zhou2022, Fuda2024, Biller2024, Chen2025}}, the light curve of Ross~458C shows a stable variability amplitude and shape {in two epochs separated} by 7.5~yrs or more than 5600 rotations, {using our newly derived} rotational period of {11.57$\pm$0.09~hr}. 
{Apart from Ross~458C},  2M2228--4310 is a fast-rotator (P$\sim$1.43~hr) T6.5 dwarf that shows {light curve} stability between different epochs \citep{Buenzli2012, Yang2016}.  
2M2228--4310 shows {pulsating} radio emissions with a tentative period of half the rotational period of 2M2228--4310 \citep{Wandia2026}, suggesting its stable variability might be {modulated by the presence of a magnetic field.}

{Ross~458C has no measurements of radio-emission, and it has a rotational} period an order of magnitude longer than 2M2228--4310, which decreases the likelihood that Ross~458C has a strong magnetic field. A measurement of the radio emission of Ross~458C would clarify if the stable light curve of the object is most likely due to {magnetospheric} emission or instead due to very stable weather patterns. 

In addition, further analysis of the JWST/NIRSpec light curves in longer wavelengths and retrieval analysis of the spectra would conclude if disequilibrium chemistry is present, potentially indicating the existence of thermochemical instabilities that could also introduce spectroscopic variability as for earlier spectral type brown dwarfs \citep{Oliveros-Gomez2026}.

\section{Conclusions}\label{sec:conclusions}

\begin{enumerate}
    \item {We discovered that the light curve of Ross~458C is consistent in variability amplitude and shape in two epochs separated more than 7.5~yrs.}

    \item We measured a rotational period of {11.57$\pm$0.09~hr}, in discrepancy to the $\sim$6.75~hr period measured by \cite{Manjavacas2019b} in the white, the $J$-band and $H$-band light curves. 
    
    \item We concluded that the rotational period measured by \cite{Manjavacas2019b} is a second harmonic of the actual rotational period. {This likely appears due to the two slightly different height peaks per rotational period in the JWST/NIRSpec light curve, undistinguishable with the higher uncertainties, {low cadence and relatively short time coverage of the HST/WFC3 light curve data}}. 

    \item We measured a variability amplitude of 1.47$\pm$0.19\% in the white light curve, 1.48$\pm$0.23\% in the $J$-band light curve, 1.60$\pm$0.49\% in the $H$-band light curve and 1.06$\pm$0.32\% in the water band light curve, consistent with the variability amplitude measured in the HST/WFC3 light curve.

    \item We measured {similar} ratios between the maximum and minimum spectra for the JWST/NIRSpec light curves and the HST/WFC3 light curves, further supporting that the light curve of Ross~458C is similar between the two epochs.

    \item {When BOTS observations must be split into multiple visits due to data volume limits, we recommend executing WATA at the start of each visit. This ensures identical target centering across both executions, significantly reducing instrumental systematics and baseline shifts between the resulting light curves.}
    
\end{enumerate}


\begin{acknowledgments}

This work is based on observations made with the NASA/ESA/CSA James Webb Space Telescope. The data were obtained from the Mikulski Archive for Space Telescopes at the Space Telescope Science Institute, which is operated by the Association of Universities for Research in Astronomy, Inc., under NASA contract NAS 5-03127 for JWST. These observations are associated with program GO~5226.

Support for program GO~5226 was provided by NASA through a grant from the Space Telescope Science Institute, which is operated by the Association of Universities for Research in Astronomy, Inc., under NASA contract NAS 5-03127.

{The JWST data presented in this article were obtained from the Mikulski Archive for Space Telescopes (MAST) at the Space Telescope Science Institute. The specific observations analyzed can be accessed via \dataset[doi: 10.17909/8gec-1805]{https://doi.org/10.17909/8gec-1805}.}

Paulo Miles-P\'aez acknowledges the use of the grant RYC2021-031173-I funded by MCIN/AEI/10.13039/501100011033 and by the “European Union NextGenerationEU/PRTR” as well as project PID2022-137241NB-C42 funded by
the Spanish “Ministerio de Ciencia, Innovaciion y Universidades”.

This work is partly based on data from the EMIR instrument, the near-infrared camera and spectrograph mounted on the Gran Telescopio Canarias.

Zhoujian Zhang acknowledges the support from program HST-AR-18150.001-A through a grant from the STScI under NASA contract NAS5-26555.

\end{acknowledgments}

%

\vspace{5mm}
\facilities{James Webb Space Telescope/NIRSpec}


\software{Astropy \citep{astropy:2013, astropy:2018, astropy:2022}}

\bibliography{Ross458c_HST}{}

@ARTICLE{Buenzli2012,
       author = {{Buenzli}, Esther and {Apai}, D{\'a}niel and {Morley}, Caroline V. and {Flateau}, Davin and {Showman}, Adam P. and {Burrows}, Adam and {Marley}, Mark S. and {Lewis}, Nikole K. and {Reid}, I. Neill},
        title = "{Vertical Atmospheric Structure in a Variable Brown Dwarf: Pressure-dependent Phase Shifts in Simultaneous Hubble Space Telescope-Spitzer Light Curves}",
      journal = {\apjl},
         year = 2012,
        month = dec,
       volume = {760},
       number = {2},
          eid = {L31},
        pages = {L31},
          doi = {10.1088/2041-8205/760/2/L31},
archivePrefix = {arXiv},
       eprint = {1210.6654},
 primaryClass = {astro-ph.SR},
       adsurl = {https://ui.adsabs.harvard.edu/abs/2012ApJ...760L..31B}
}

@ARTICLE{Wandia2026,
       author = {{Wandia}, Kelvin and {Garrett}, Michael A. and {Golden}, Aaron and {Hallinan}, Gregg and {Williams-Baldwin}, David and {Lucatelli}, Geferson and {Beswick}, Robert J. and {Radcliffe}, Jack F. and {Siemion}, Andrew and {Myburgh}, Talon},
        title = "{Radio activity from the rapidly rotating T dwarf 2MASS 2228-4310}",
      journal = {\mnras},
         year = 2026,
        month = mar,
       volume = {546},
       number = {4},
          eid = {stag085},
        pages = {stag085},
          doi = {10.1093/mnras/stag085},
archivePrefix = {arXiv},
       eprint = {2601.04158},
 primaryClass = {astro-ph.SR},
       adsurl = {https://ui.adsabs.harvard.edu/abs/2026MNRAS.546ag085W}
}

@ARTICLE{Yang2016,
       author = {{Yang}, Hao and {Apai}, D{\'a}niel and {Marley}, Mark S. and {Karalidi}, Theodora and {Flateau}, Davin and {Showman}, Adam P. and {Metchev}, Stanimir and {Buenzli}, Esther and {Radigan}, Jacqueline and {Artigau}, {\'E}tienne and {Lowrance}, Patrick J. and {Burgasser}, Adam J.},
        title = "{Extrasolar Storms: Pressure-dependent Changes in Light-curve Phase in Brown Dwarfs from Simultaneous HST and Spitzer Observations}",
      journal = {\apj},
         year = 2016,
        month = jul,
       volume = {826},
       number = {1},
          eid = {8},
        pages = {8},
          doi = {10.3847/0004-637X/826/1/8},
archivePrefix = {arXiv},
       eprint = {1605.02708},
 primaryClass = {astro-ph.EP},
       adsurl = {https://ui.adsabs.harvard.edu/abs/2016ApJ...826....8Y}
}

@software{jwst_pipeline,
       author = {{Bushouse}, Howard and {Eisenhamer}, Jonathan and {Dencheva}, Nadia and {Davies}, James and {Greenfield}, Perry and {Morrison}, Jane and {Hodge}, Phil and {Simon}, Bernie and {Grumm}, David and {Droettboom}, Michael and {Slavich}, Edward and {Sosey}, Megan and {Pauly}, Tyler and {Miller}, Todd and {Jedrzejewski}, Robert and {Hack}, Warren and {Davis}, David and {Crawford}, Steven and {Law}, David and {Gordon}, Karl and {Regan}, Michael and {Cara}, Mihai and {MacDonald}, Ken and {Bradley}, Larry and {Shanahan}, Clare and {Jamieson}, William and {Teodoro}, Mairan and {Williams}, Thomas and {Pena-Guerrero}, Maria and {Graham}, Brett and {Molter}, Edward and {Brandt}, Timothy and {Hayes}, Christian and {Cooper}, Rachel and {Clarke}, Melanie and {Filippazzo}, Joseph},
        title = "{JWST Calibration Pipeline}",
         year = 2025,
        month = apr,
          eid = {10.5281/zenodo.15178003},
          doi = {10.5281/zenodo.15178003},
      version = {1.18.0},
    publisher = {Zenodo},
       adsurl = {https://ui.adsabs.harvard.edu/abs/2025zndo..15178003B}
}

@ARTICLE{Horne_Baliunas1986,
       author = {{Horne}, J.~H. and {Baliunas}, S.~L.},
        title = "{A Prescription for Period Analysis of Unevenly Sampled Time Series}",
      journal = {\apj},
         year = 1986,
        month = mar,
       volume = {302},
        pages = {757},
          doi = {10.1086/164037},
       adsurl = {https://ui.adsabs.harvard.edu/abs/1986ApJ...302..757H}
}

@article{celerite2,
   author = {{Foreman-Mackey}, D.},
    title = "{Scalable Backpropagation for Gaussian Processes using Celerite}",
  journal = {Research Notes of the American Astronomical Society},
     year = 2018,
    month = feb,
   volume = 2,
   number = 1,
    pages = {31},
      doi = {10.3847/2515-5172/aaaf6c},
   adsurl = {http://adsabs.harvard.edu/abs/2018RNAAS...2a..31F}
}

@ARTICLE{Mortier2015,
       author = {{Mortier}, A. and {Faria}, J.~P. and {Correia}, C.~M. and {Santerne}, A. and {Santos}, N.~C.},
        title = "{BGLS: A Bayesian formalism for the generalised Lomb-Scargle periodogram}",
      journal = {\aap},
         year = 2015,
        month = jan,
       volume = {573},
          eid = {A101},
        pages = {A101},
          doi = {10.1051/0004-6361/201424908},
archivePrefix = {arXiv},
       eprint = {1412.0467},
 primaryClass = {astro-ph.IM},
       adsurl = {https://ui.adsabs.harvard.edu/abs/2015A&A...573A.101M}
}

@ARTICLE{Baluev2008,
       author = {{Baluev}, R.~V.},
        title = "{Assessing the statistical significance of periodogram peaks}",
      journal = {\mnras},
         year = 2008,
        month = apr,
       volume = {385},
       number = {3},
        pages = {1279-1285},
          doi = {10.1111/j.1365-2966.2008.12689.x},
archivePrefix = {arXiv},
       eprint = {0711.0330},
 primaryClass = {astro-ph},
       adsurl = {https://ui.adsabs.harvard.edu/abs/2008MNRAS.385.1279B}
}

@ARTICLE{Meynardie2025,
       author = {{Meynardie}, William W. and {Meyer}, Michael R. and {MacDonald}, Ryan J. and {Calissendorff}, Per and {Mullens}, Elijah and {Zarazua}, Gabriel Munoz and {Roy}, Anuranj and {Ganta}, Hansica and {Gonzales}, Eileen C. and {Adams}, Arthur and {Lewis}, Nikole and {Hong}, Yucian and {Lunine}, Jonathan},
        title = "{Ross 458 C: Gas Giant or Brown Dwarf?}",
      journal = {\apj},
         year = 2025,
        month = dec,
       volume = {994},
       number = {2},
          eid = {237},
        pages = {237},
          doi = {10.3847/1538-4357/ae0ad0},
archivePrefix = {arXiv},
       eprint = {2509.22803},
 primaryClass = {astro-ph.EP},
       adsurl = {https://ui.adsabs.harvard.edu/abs/2025ApJ...994..237M}
}

@ARTICLE{Gaarn2023,
       author = {{Gaarn}, Josefine and {Burningham}, Ben and {Faherty}, Jacqueline K. and {Visscher}, Channon and {Marley}, Mark S. and {Gonzales}, Eileen C. and {Calamari}, Emily and {Bardalez Gagliuffi}, Daniella and {Lupu}, Roxana and {Freedman}, Richard},
        title = "{The puzzle of the formation of T8 dwarf Ross 458c}",
      journal = {\mnras},
         year = 2023,
        month = jun,
       volume = {521},
       number = {4},
        pages = {5761-5775},
          doi = {10.1093/mnras/stad753},
archivePrefix = {arXiv},
       eprint = {2303.16863},
 primaryClass = {astro-ph.SR},
       adsurl = {https://ui.adsabs.harvard.edu/abs/2023MNRAS.521.5761G}
}

@ARTICLE{Morley2012,
       author = {{Morley}, Caroline V. and {Fortney}, Jonathan J. and {Marley}, Mark S. and {Visscher}, Channon and {Saumon}, Didier and {Leggett}, S.~K.},
        title = "{Neglected Clouds in T and Y Dwarf Atmospheres}",
      journal = {\apj},
         year = 2012,
        month = sep,
       volume = {756},
       number = {2},
          eid = {172},
        pages = {172},
          doi = {10.1088/0004-637X/756/2/172},
archivePrefix = {arXiv},
       eprint = {1206.4313},
 primaryClass = {astro-ph.SR},
       adsurl = {https://ui.adsabs.harvard.edu/abs/2012ApJ...756..172M}
}

@ARTICLE{Burningham2010,
       author = {{Burningham}, Ben and {Leggett}, S.~K. and {Homeier}, D. and {Saumon}, D. and {Lucas}, P.~W. and {Pinfield}, D.~J. and {Tinney}, C.~G. and {Allard}, F. and {Marley}, M.~S. and {Jones}, H.~R.~A. and {Murray}, D.~N. and {Ishii}, M. and {Day-Jones}, A. and {Gomes}, J. and {Zhang}, Z.~H.},
        title = "{The properties of the T8.5p dwarf Ross 458C}",
      journal = {\mnras},
         year = 2011,
        month = jul,
       volume = {414},
       number = {4},
        pages = {3590-3598},
          doi = {10.1111/j.1365-2966.2011.18664.x},
archivePrefix = {arXiv},
       eprint = {1103.1617},
 primaryClass = {astro-ph.SR},
       adsurl = {https://ui.adsabs.harvard.edu/abs/2011MNRAS.414.3590B}
}

@ARTICLE{Magazzu1993,
       author = {{Magazzu}, Antonio and {Martin}, Eduardo L. and {Rebolo}, Rafael},
        title = "{A Spectroscopic Test for Substellar Objects}",
      journal = {\apjl},
         year = 1993,
        month = feb,
       volume = {404},
        pages = {L17},
          doi = {10.1086/186733},
       adsurl = {https://ui.adsabs.harvard.edu/abs/1993ApJ...404L..17M}
}

@ARTICLE{Chabrier1996,
       author = {{Chabrier}, Gilles and {Baraffe}, Isabelle and {Plez}, Bertrand},
        title = "{Mass-Luminosity Relationship and Lithium Depletion for Very Low Mass Stars}",
      journal = {\apjl},
         year = 1996,
        month = mar,
       volume = {459},
        pages = {L91},
          doi = {10.1086/309951},
       adsurl = {https://ui.adsabs.harvard.edu/abs/1996ApJ...459L..91C}
}

@ARTICLE{Burgasser2010_Ross,
       author = {{Burgasser}, Adam J. and {Simcoe}, Robert A. and {Bochanski}, John J. and {Saumon}, Didier and {Mamajek}, Eric E. and {Cushing}, Michael C. and {Marley}, Mark S. and {McMurtry}, Craig and {Pipher}, Judith L. and {Forrest}, William J.},
        title = "{Clouds in the Coldest Brown Dwarfs: Fire Spectroscopy of Ross 458C}",
      journal = {\apj},
         year = 2010,
        month = dec,
       volume = {725},
       number = {2},
        pages = {1405-1420},
          doi = {10.1088/0004-637X/725/2/1405},
archivePrefix = {arXiv},
       eprint = {1009.5722},
 primaryClass = {astro-ph.SR},
       adsurl = {https://ui.adsabs.harvard.edu/abs/2010ApJ...725.1405B}
}

@ARTICLE{West2008,
       author = {{West}, Andrew A. and {Hawley}, Suzanne L. and {Bochanski}, John J. and {Covey}, Kevin R. and {Reid}, I. Neill and {Dhital}, Saurav and {Hilton}, Eric J. and {Masuda}, Michael},
        title = "{Constraining the Age-Activity Relation for Cool Stars: The Sloan Digital Sky Survey Data Release 5 Low-Mass Star Spectroscopic Sample}",
      journal = {\aj},
         year = 2008,
        month = mar,
       volume = {135},
       number = {3},
        pages = {785-795},
          doi = {10.1088/0004-6256/135/3/785},
archivePrefix = {arXiv},
       eprint = {0712.1590},
 primaryClass = {astro-ph},
       adsurl = {https://ui.adsabs.harvard.edu/abs/2008AJ....135..785W}
}

@ARTICLE{Rustamkulov2022,
       author = {{Rustamkulov}, Zafar and {Sing}, David K. and {Liu}, Rongrong and {Wang}, Ashley},
        title = "{Analysis of a JWST NIRSpec Lab Time Series: Characterizing Systematics, Recovering Exoplanet Transit Spectroscopy, and Constraining a Noise Floor}",
      journal = {\apjl},
         year = 2022,
        month = mar,
       volume = {928},
       number = {1},
          eid = {L7},
        pages = {L7},
          doi = {10.3847/2041-8213/ac5b6f},
archivePrefix = {arXiv},
       eprint = {2203.04173},
 primaryClass = {astro-ph.EP},
       adsurl = {https://ui.adsabs.harvard.edu/abs/2022ApJ...928L...7R}
}

@ARTICLE{Alam2025,
       author = {{Alam}, Munazza K. and {Gao}, Peter and {Adams Redai}, Jea and {Wallack}, Nicole L. and {Wogan}, Nicholas F. and {Aguichine}, Artyom and {Dattilo}, Anne and {Alderson}, Lili and {Batalha}, Natasha E. and {Batalha}, Natalie M. and {Kirk}, James and {L{\'o}pez-Morales}, Mercedes and {Meech}, Annabella and {Moran}, Sarah E. and {Teske}, Johanna and {Wakeford}, Hannah R. and {Wolfgang}, Angie},
        title = "{JWST COMPASS: The First Near- to Mid-infrared Transmission Spectrum of the Hot Super-Earth L 168-9 b}",
      journal = {\aj},
         year = 2025,
        month = jan,
       volume = {169},
       number = {1},
          eid = {15},
        pages = {15},
          doi = {10.3847/1538-3881/ad8eb5},
archivePrefix = {arXiv},
       eprint = {2411.03154},
 primaryClass = {astro-ph.EP},
       adsurl = {https://ui.adsabs.harvard.edu/abs/2025AJ....169...15A}
}

@ARTICLE{Manjavacas2019b,
       author = {{Manjavacas}, Elena and {Apai}, D{\'a}niel and {Lew}, Ben W.~P. and {Zhou}, Yifan and {Schneider}, Glenn and {Burgasser}, Adam J. and {Karalidi}, Theodora and {Miles-P{\'a}ez}, Paulo A. and {Lowrance}, Patrick J. and {Cowan}, Nicolas and {Bedin}, Luigi R. and {Marley}, Mark S. and {Metchev}, Stan and {Radigan}, Jacqueline},
        title = "{Cloud Atlas: Rotational Spectral Modulations and Potential Sulfide Clouds in the Planetary-mass, Late T-type Companion Ross 458C}",
      journal = {\apjl},
         year = 2019,
        month = apr,
       volume = {875},
       number = {2},
          eid = {L15},
        pages = {L15},
          doi = {10.3847/2041-8213/ab13b9},
archivePrefix = {arXiv},
       eprint = {1903.10702},
 primaryClass = {astro-ph.SR},
       adsurl = {https://ui.adsabs.harvard.edu/abs/2019ApJ...875L..15M}
}

@ARTICLE{Goldman2010,
       author = {{Goldman}, B. and {Marsat}, S. and {Henning}, T. and {Clemens}, C. and {Greiner}, J.},
        title = "{A new benchmark T8-9 brown dwarf and a couple of new mid-T dwarfs from the UKIDSS DR5+ LAS}",
      journal = {\mnras},
         year = 2010,
        month = jun,
       volume = {405},
       number = {2},
        pages = {1140-1152},
          doi = {10.1111/j.1365-2966.2010.16524.x},
archivePrefix = {arXiv},
       eprint = {1002.2637},
 primaryClass = {astro-ph.SR},
       adsurl = {https://ui.adsabs.harvard.edu/abs/2010MNRAS.405.1140G}
}

@ARTICLE{Jakobsen2022,
       author = {{Jakobsen}, P. and {Ferruit}, P. and {Alves de Oliveira}, C. and {Arribas}, S. and {Bagnasco}, G. and {Barho}, R. and {Beck}, T.~L. and {Birkmann}, S. and {B{\"o}ker}, T. and {Bunker}, A.~J. and {Charlot}, S. and {de Jong}, P. and {de Marchi}, G. and {Ehrenwinkler}, R. and {Falcolini}, M. and {Fels}, R. and {Franx}, M. and {Franz}, D. and {Funke}, M. and {Giardino}, G. and {Gnata}, X. and {Holota}, W. and {Honnen}, K. and {Jensen}, P.~L. and {Jentsch}, M. and {Johnson}, T. and {Jollet}, D. and {Karl}, H. and {Kling}, G. and {K{\"o}hler}, J. and {Kolm}, M. -G. and {Kumari}, N. and {Lander}, M.~E. and {Lemke}, R. and {L{\'o}pez-Caniego}, M. and {L{\"u}tzgendorf}, N. and {Maiolino}, R. and {Manjavacas}, E. and {Marston}, A. and {Maschmann}, M. and {Maurer}, R. and {Messerschmidt}, B. and {Moseley}, S.~H. and {Mosner}, P. and {Mott}, D.~B. and {Muzerolle}, J. and {Pirzkal}, N. and {Pittet}, J. -F. and {Plitzke}, A. and {Posselt}, W. and {Rapp}, B. and {Rauscher}, B.~J. and {Rawle}, T. and {Rix}, H. -W. and {R{\"o}del}, A. and {Rumler}, P. and {Sabbi}, E. and {Salvignol}, J. -C. and {Schmid}, T. and {Sirianni}, M. and {Smith}, C. and {Strada}, P. and {te Plate}, M. and {Valenti}, J. and {Wettemann}, T. and {Wiehe}, T. and {Wiesmayer}, M. and {Willott}, C.~J. and {Wright}, R. and {Zeidler}, P. and {Zincke}, C.},
        title = "{The Near-Infrared Spectrograph (NIRSpec) on the James Webb Space Telescope. I. Overview of the instrument and its capabilities}",
      journal = {\aap},
         year = 2022,
        month = may,
       volume = {661},
          eid = {A80},
        pages = {A80},
          doi = {10.1051/0004-6361/202142663},
archivePrefix = {arXiv},
       eprint = {2202.03305},
 primaryClass = {astro-ph.IM},
       adsurl = {https://ui.adsabs.harvard.edu/abs/2022A&A...661A..80J}
}

@article{astropy:2013,
        Adsurl = {http://adsabs.harvard.edu/abs/2013A%26A...558A..33A},
        Archiveprefix = {arXiv},
        Author = {{Astropy Collaboration} and {Robitaille}, T.~P. and {Tollerud}, E.~J. and {Greenfield}, P. and {Droettboom}, M. and {Bray}, E. and {Aldcroft}, T. and {Davis}, M. and {Ginsburg}, A. and {Price-Whelan}, A.~M. and {Kerzendorf}, W.~E. and {Conley}, A. and {Crighton}, N. and {Barbary}, K. and {Muna}, D. and {Ferguson}, H. and {Grollier}, F. and {Parikh}, M.~M. and {Nair}, P.~H. and {Unther}, H.~M. and {Deil}, C. and {Woillez}, J. and {Conseil}, S. and {Kramer}, R. and {Turner}, J.~E.~H. and {Singer}, L. and {Fox}, R. and {Weaver}, B.~A. and {Zabalza}, V. and {Edwards}, Z.~I. and {Azalee Bostroem}, K. and {Burke}, D.~J. and {Casey}, A.~R. and {Crawford}, S.~M. and {Dencheva}, N. and {Ely}, J. and {Jenness}, T. and {Labrie}, K. and {Lim}, P.~L. and {Pierfederici}, F. and {Pontzen}, A. and {Ptak}, A. and {Refsdal}, B. and {Servillat}, M. and {Streicher}, O.},
        Doi = {10.1051/0004-6361/201322068},
        Eid = {A33},
        Eprint = {1307.6212},
        Journal = {\aap},
        Month = oct,
        Pages = {A33},
        Primaryclass = {astro-ph.IM},
        Title = {{Astropy: A community Python package for astronomy}},
        Volume = 558,
        Year = 2013}

@ARTICLE{astropy:2018,
               author = {{Astropy Collaboration} and {Price-Whelan}, A.~M. and
                 {Sip{\H{o}}cz}, B.~M. and {G{\"u}nther}, H.~M. and {Lim}, P.~L. and
                 {Crawford}, S.~M. and {Conseil}, S. and {Shupe}, D.~L. and
                 {Craig}, M.~W. and {Dencheva}, N. and {Ginsburg}, A. and {Vand
                erPlas}, J.~T. and {Bradley}, L.~D. and {P{\'e}rez-Su{\'a}rez}, D. and
                 {de Val-Borro}, M. and {Aldcroft}, T.~L. and {Cruz}, K.~L. and
                 {Robitaille}, T.~P. and {Tollerud}, E.~J. and {Ardelean}, C. and
                 {Babej}, T. and {Bach}, Y.~P. and {Bachetti}, M. and {Bakanov}, A.~V. and
                 {Bamford}, S.~P. and {Barentsen}, G. and {Barmby}, P. and
                 {Baumbach}, A. and {Berry}, K.~L. and {Biscani}, F. and {Boquien}, M. and
                 {Bostroem}, K.~A. and {Bouma}, L.~G. and {Brammer}, G.~B. and
                 {Bray}, E.~M. and {Breytenbach}, H. and {Buddelmeijer}, H. and
                 {Burke}, D.~J. and {Calderone}, G. and {Cano Rodr{\'\i}guez}, J.~L. and
                 {Cara}, M. and {Cardoso}, J.~V.~M. and {Cheedella}, S. and {Copin}, Y. and
                 {Corrales}, L. and {Crichton}, D. and {D'Avella}, D. and {Deil}, C. and
                 {Depagne}, {\'E}. and {Dietrich}, J.~P. and {Donath}, A. and
                 {Droettboom}, M. and {Earl}, N. and {Erben}, T. and {Fabbro}, S. and
                 {Ferreira}, L.~A. and {Finethy}, T. and {Fox}, R.~T. and
                 {Garrison}, L.~H. and {Gibbons}, S.~L.~J. and {Goldstein}, D.~A. and
                 {Gommers}, R. and {Greco}, J.~P. and {Greenfield}, P. and
                 {Groener}, A.~M. and {Grollier}, F. and {Hagen}, A. and {Hirst}, P. and
                 {Homeier}, D. and {Horton}, A.~J. and {Hosseinzadeh}, G. and {Hu}, L. and
                 {Hunkeler}, J.~S. and {Ivezi{\'c}}, {\v{Z}}. and {Jain}, A. and
                 {Jenness}, T. and {Kanarek}, G. and {Kendrew}, S. and {Kern}, N.~S. and
                 {Kerzendorf}, W.~E. and {Khvalko}, A. and {King}, J. and {Kirkby}, D. and
                 {Kulkarni}, A.~M. and {Kumar}, A. and {Lee}, A. and {Lenz}, D. and
                 {Littlefair}, S.~P. and {Ma}, Z. and {Macleod}, D.~M. and
                 {Mastropietro}, M. and {McCully}, C. and {Montagnac}, S. and
                 {Morris}, B.~M. and {Mueller}, M. and {Mumford}, S.~J. and {Muna}, D. and
                 {Murphy}, N.~A. and {Nelson}, S. and {Nguyen}, G.~H. and
                 {Ninan}, J.~P. and {N{\"o}the}, M. and {Ogaz}, S. and {Oh}, S. and
                 {Parejko}, J.~K. and {Parley}, N. and {Pascual}, S. and {Patil}, R. and
                 {Patil}, A.~A. and {Plunkett}, A.~L. and {Prochaska}, J.~X. and
                 {Rastogi}, T. and {Reddy Janga}, V. and {Sabater}, J. and
                 {Sakurikar}, P. and {Seifert}, M. and {Sherbert}, L.~E. and
                 {Sherwood-Taylor}, H. and {Shih}, A.~Y. and {Sick}, J. and
                 {Silbiger}, M.~T. and {Singanamalla}, S. and {Singer}, L.~P. and
                 {Sladen}, P.~H. and {Sooley}, K.~A. and {Sornarajah}, S. and
                 {Streicher}, O. and {Teuben}, P. and {Thomas}, S.~W. and
                 {Tremblay}, G.~R. and {Turner}, J.~E.~H. and {Terr{\'o}n}, V. and
                 {van Kerkwijk}, M.~H. and {de la Vega}, A. and {Watkins}, L.~L. and
                 {Weaver}, B.~A. and {Whitmore}, J.~B. and {Woillez}, J. and
                 {Zabalza}, V. and {Astropy Contributors}},
                title = "{The Astropy Project: Building an Open-science Project and Status of the v2.0 Core Package}",
              journal = {\aj},
                 year = 2018,
                month = sep,
               volume = {156},
               number = {3},
                  eid = {123},
                pages = {123},
                  doi = {10.3847/1538-3881/aabc4f},
        archivePrefix = {arXiv},
               eprint = {1801.02634},
         primaryClass = {astro-ph.IM},
               adsurl = {https://ui.adsabs.harvard.edu/abs/2018AJ....156..123A}
        }

@article{KassRaftery1995,
  author  = {Kass, Robert E. and Raftery, Adrian E.},
  title   = {Bayes Factors},
  journal = {Journal of the American Statistical Association},
  year    = {1995},
  volume  = {90},
  number  = {430},
  pages   = {773--795},
  doi     = {10.1080/01621459.1995.10476572}
}

@ARTICLE{Wright2023,
       author = {{Wright}, Gillian S. and {Rieke}, George H. and {Glasse}, Alistair and {Ressler}, Michael and {Garc{\'\i}a Mar{\'\i}n}, Macarena and {Aguilar}, Jonathan and {Alberts}, Stacey and {{\'A}lvarez-M{\'a}rquez}, Javier and {Argyriou}, Ioannis and {Banks}, Kimberly and {Baudoz}, Pierre and {Boccaletti}, Anthony and {Bouchet}, Patrice and {Bouwman}, Jeroen and {Brandl}, Bernard R. and {Breda}, David and {Bright}, Stacey and {Cale}, Steven and {Colina}, Luis and {Cossou}, Christophe and {Coulais}, Alain and {Cracraft}, Misty and {De Meester}, Wim and {Dicken}, Daniel and {Engesser}, Michael and {Etxaluze}, Mireya and {Fox}, Ori D. and {Friedman}, Scott and {Fu}, Henry and {Gasman}, Danny and {G{\'a}sp{\'a}r}, Andr{\'a}s and {Gastaud}, Ren{\'e} and {Geers}, Vincent and {Glauser}, Adrian Michael and {Gordon}, Karl D. and {Greene}, Thomas and {Greve}, Thomas R. and {Grundy}, Timothy and {G{\"u}del}, Manuel and {Guillard}, Pierre and {Haderlein}, Peter and {Hashimoto}, Ryan and {Henning}, Thomas and {Hines}, Dean and {Holler}, Bryan and {Detre}, {\"O}rs Hunor and {Jahromi}, Amir and {James}, Bryan and {Jones}, Olivia C. and {Justtanont}, Kay and {Kavanagh}, Patrick and {Kendrew}, Sarah and {Klaassen}, Pamela and {Krause}, Oliver and {Labiano}, Alvaro and {Lagage}, Pierre-Olivier and {Lambros}, Scott and {Larson}, Kirsten and {Law}, David and {Lee}, David and {Libralato}, Mattia and {Lorenzo Alverez}, Jose and {Meixner}, Margaret and {Morrison}, Jane and {Mueller}, Migo and {Murray}, Katherine and {Mycroft}, Matthew and {Myers}, Richard and {Nayak}, Omnarayani and {Naylor}, Bret and {Nickson}, Bryony and {Noriega-Crespo}, Alberto and {{\"O}stlin}, G{\"o}ran and {O'Sullivan}, Brian and {Ottens}, Richard and {Patapis}, Polychronis and {Penanen}, Konstantin and {Pietraszkiewicz}, Martin and {Ray}, Tom and {Regan}, Michael and {Roteliuk}, Anthony and {Royer}, Pierre and {Samara-Ratna}, Piyal and {Samuelson}, Bridget and {Sargent}, Beth A. and {Scheithauer}, Silvia and {Schneider}, Analyn and {Schreiber}, J{\"u}rgen and {Shaughnessy}, Bryan and {Sheehan}, Evan and {Shivaei}, Irene and {Sloan}, G.~C. and {Tamas}, Laszlo and {Teague}, Kelly and {Temim}, Tea and {Tikkanen}, Tuomo and {Tustain}, Samuel and {van Dishoeck}, Ewine F. and {Vandenbussche}, Bart and {Weilert}, Mark and {Whitehouse}, Paul and {Wolff}, Schuyler},
        title = "{The Mid-infrared Instrument for JWST and Its In-flight Performance}",
      journal = {\pasp},
         year = 2023,
        month = apr,
       volume = {135},
       number = {1046},
          eid = {048003},
        pages = {048003},
          doi = {10.1088/1538-3873/acbe66},
       adsurl = {https://ui.adsabs.harvard.edu/abs/2023PASP..135d8003W}
}

@ARTICLE{Bocker2023,
       author = {{B{\"o}ker}, T. and {Beck}, T.~L. and {Birkmann}, S.~M. and {Giardino}, G. and {Keyes}, C. and {Kumari}, N. and {Muzerolle}, J. and {Rawle}, T. and {Zeidler}, P. and {Abul-Huda}, Y. and {Alves de Oliveira}, C. and {Arribas}, S. and {Bechtold}, K. and {Bhatawdekar}, R. and {Bonaventura}, N. and {Bunker}, A.~J. and {Cameron}, A.~J. and {Carniani}, S. and {Charlot}, S. and {Curti}, M. and {Espinoza}, N. and {Ferruit}, P. and {Franx}, M. and {Jakobsen}, P. and {Karakla}, D. and {L{\'o}pez-Caniego}, M. and {L{\"u}tzgendorf}, N. and {Maiolino}, R. and {Manjavacas}, E. and {Marston}, A.~P. and {Moseley}, S.~H. and {Ogle}, P. and {Perna}, M. and {Pe{\~n}a-Guerrero}, M. and {Pirzkal}, N. and {Plesha}, R. and {Proffitt}, C.~R. and {Rauscher}, B.~J. and {Rix}, H.-W. and {Rodr{\'\i}guez del Pino}, B. and {Rustamkulov}, Z. and {Sabbi}, E. and {Sing}, D.~K. and {Sirianni}, M. and {te Plate}, M. and {{\'U}beda}, L. and {Wahlgren}, G.~M. and {Wislowski}, E. and {Wu}, R. and {Willott}, Chris J.},
        title = "{In-orbit Performance of the Near-infrared Spectrograph NIRSpec on the James Webb Space Telescope}",
      journal = {\pasp},
         year = 2023,
        month = mar,
       volume = {135},
       number = {1045},
          eid = {038001},
        pages = {038001},
          doi = {10.1088/1538-3873/acb846},
archivePrefix = {arXiv},
       eprint = {2301.13766},
 primaryClass = {astro-ph.IM},
       adsurl = {https://ui.adsabs.harvard.edu/abs/2023PASP..135c8001B}
}

@ARTICLE{Rigby2023,
       author = {{Rigby}, Jane and {Perrin}, Marshall and {McElwain}, Michael and {Kimble}, Randy and {Friedman}, Scott and {Lallo}, Matt and {Doyon}, Ren{\'e} and {Feinberg}, Lee and {Ferruit}, Pierre and {Glasse}, Alistair and {Rieke}, Marcia and {Rieke}, George and {Wright}, Gillian and {Willott}, Chris and {Colon}, Knicole and {Milam}, Stefanie and {Neff}, Susan and {Stark}, Christopher and {Valenti}, Jeff and {Abell}, Jim and {Abney}, Faith and {Abul-Huda}, Yasin and {Acton}, D. Scott and {Adams}, Evan and {Adler}, David and {Aguilar}, Jonathan and {Ahmed}, Nasif and {Albert}, Lo{\"\i}c and {Alberts}, Stacey and {Aldridge}, David and {Allen}, Marsha and {Altenburg}, Martin and {{\'A}lvarez-M{\'a}rquez}, Javier and {Alves de Oliveira}, Catarina and {Andersen}, Greg and {Anderson}, Harry and {Anderson}, Sara and {Argyriou}, Ioannis and {Armstrong}, Amber and {Arribas}, Santiago and {Artigau}, Etienne and {Arvai}, Amanda and {Atkinson}, Charles and {Bacon}, Gregory and {Bair}, Thomas and {Banks}, Kimberly and {Barrientes}, Jaclyn and {Barringer}, Bruce and {Bartosik}, Peter and {Bast}, William and {Baudoz}, Pierre and {Beatty}, Thomas and {Bechtold}, Katie and {Beck}, Tracy and {Bergeron}, Eddie and {Bergkoetter}, Matthew and {Bhatawdekar}, Rachana and {Birkmann}, Stephan and {Blazek}, Ronald and {Blome}, Claire and {Boccaletti}, Anthony and {B{\"o}ker}, Torsten and {Boia}, John and {Bonaventura}, Nina and {Bond}, Nicholas and {Bosley}, Kari and {Boucarut}, Ray and {Bourque}, Matthew and {Bouwman}, Jeroen and {Bower}, Gary and {Bowers}, Charles and {Boyer}, Martha and {Bradley}, Larry and {Brady}, Greg and {Braun}, Hannah and {Breda}, David and {Bresnahan}, Pamela and {Bright}, Stacey and {Britt}, Christopher and {Bromenschenkel}, Asa and {Brooks}, Brian and {Brooks}, Keira and {Brown}, Bob and {Brown}, Matthew and {Brown}, Patricia and {Bunker}, Andy and {Burger}, Matthew and {Bushouse}, Howard and {Cale}, Steven and {Cameron}, Alex and {Cameron}, Peter and {Canipe}, Alicia and {Caplinger}, James and {Caputo}, Francis and {Cara}, Mihai and {Carey}, Larkin and {Carniani}, Stefano and {Carrasquilla}, Maria and {Carruthers}, Margaret and {Case}, Michael and {Catherine}, Riggs and {Chance}, Don and {Chapman}, George and {Charlot}, St{\'e}phane and {Charlow}, Brian and {Chayer}, Pierre and {Chen}, Bin and {Cherinka}, Brian and {Chichester}, Sarah and {Chilton}, Zack and {Chonis}, Taylor and {Clampin}, Mark and {Clark}, Charles and {Clark}, Kerry and {Coe}, Dan and {Coleman}, Benee and {Comber}, Brian and {Comeau}, Tom and {Connolly}, Dennis and {Cooper}, James and {Cooper}, Rachel and {Coppock}, Eric and {Correnti}, Matteo and {Cossou}, Christophe and {Coulais}, Alain and {Coyle}, Laura and {Cracraft}, Misty and {Curti}, Mirko and {Cuturic}, Steven and {Davis}, Katherine and {Davis}, Michael and {Dean}, Bruce and {DeLisa}, Amy and {deMeester}, Wim and {Dencheva}, Nadia and {Dencheva}, Nadezhda and {DePasquale}, Joseph and {Deschenes}, Jeremy and {Hunor Detre}, {\"O}rs and {Diaz}, Rosa and {Dicken}, Dan and {DiFelice}, Audrey and {Dillman}, Matthew and {Dixon}, William and {Doggett}, Jesse and {Donaldson}, Tom and {Douglas}, Rob and {DuPrie}, Kimberly and {Dupuis}, Jean and {Durning}, John and {Easmin}, Nilufar and {Eck}, Weston and {Edeani}, Chinwe and {Egami}, Eiichi and {Ehrenwinkler}, Ralf and {Eisenhamer}, Jonathan and {Eisenhower}, Michael and {Elie}, Michelle and {Elliott}, James and {Elliott}, Kyle and {Ellis}, Tracy and {Engesser}, Michael and {Espinoza}, Nestor and {Etienne}, Odessa and {Etxaluze}, Mireya and {Falini}, Patrick and {Feeney}, Matthew and {Ferry}, Malcolm and {Filippazzo}, Joseph and {Fincham}, Brian and {Fix}, Mees and {Flagey}, Nicolas and {Florian}, Michael and {Flynn}, Jim and {Fontanella}, Erin and {Ford}, Terrance and {Forshay}, Peter and {Fox}, Ori and {Franz}, David and {Fu}, Henry and {Fullerton}, Alexander and {Galkin}, Sergey and {Galyer}, Anthony and {Garc{\'\i}a Mar{\'\i}n}, Macarena and {Gardner}, Jonathan P. and {Gardner}, Lisa and {Garland}, Dennis and {Garrett}, Bruce and {Gasman}, Danny and {Gaspar}, Andras and {Gaudreau}, Daniel and {Gauthier}, Peter and {Geers}, Vincent and {Geithner}, Paul and {Gennaro}, Mario and {Giardino}, Giovanna and {Girard}, Julien and {Giuliano}, Mark and {Glassmire}, Kirk and {Glauser}, Adrian},
        title = "{The Science Performance of JWST as Characterized in Commissioning}",
      journal = {\pasp},
         year = 2023,
        month = apr,
       volume = {135},
       number = {1046},
          eid = {048001},
        pages = {048001},
          doi = {10.1088/1538-3873/acb293},
archivePrefix = {arXiv},
       eprint = {2207.05632},
 primaryClass = {astro-ph.IM},
       adsurl = {https://ui.adsabs.harvard.edu/abs/2023PASP..135d8001R}
}

@INPROCEEDINGS{Gardner2006,
       author = {{Gardner}, Jonathan P. and {Mather}, John C. and {Clampin}, Mark and {Doyon}, Rene and {Greenhouse}, Matthew A. and {Hammel}, Heidi B. and {Hutchings}, John B. and {Jakobsen}, Peter and {Lilly}, Simon J. and {Long}, Knox S. and {Lunine}, Jonathan I. and {McCaughrean}, Mark J. and {Mountain}, Matt and {Nella}, John and {Rieke}, George H. and {Rieke}, Marcia J. and {Rix}, Hans-Walter and {Smith}, Eric P. and {Sonneborn}, George and {Stiavelli}, Massimo and {Stockman}, H.~S. and {Windhorst}, Rogier A. and {Wright}, Gillian S.},
        title = "{Science with the James Webb space telescope}",
    booktitle = {Space Telescopes and Instrumentation I: Optical, Infrared, and Millimeter},
         year = 2006,
       editor = {{Mather}, John C. and {MacEwen}, Howard A. and {de Graauw}, Mattheus W.~M.},
       series = {Society of Photo-Optical Instrumentation Engineers (SPIE) Conference Series},
       volume = {6265},
        month = jun,
          eid = {62650N},
        pages = {62650N},
          doi = {10.1117/12.670492},
       adsurl = {https://ui.adsabs.harvard.edu/abs/2006SPIE.6265E..0NG}
}

@ARTICLE{astropy:2022,
               author = {{Astropy Collaboration} and {Price-Whelan}, Adrian M. and {Lim}, Pey Lian and {Earl}, Nicholas and {Starkman}, Nathaniel and {Bradley}, Larry and {Shupe}, David L. and {Patil}, Aarya A. and {Corrales}, Lia and {Brasseur}, C.~E. and {N{"o}the}, Maximilian and {Donath}, Axel and {Tollerud}, Erik and {Morris}, Brett M. and {Ginsburg}, Adam and {Vaher}, Eero and {Weaver}, Benjamin A. and {Tocknell}, James and {Jamieson}, William and {van Kerkwijk}, Marten H. and {Robitaille}, Thomas P. and {Merry}, Bruce and {Bachetti}, Matteo and {G{"u}nther}, H. Moritz and {Aldcroft}, Thomas L. and {Alvarado-Montes}, Jaime A. and {Archibald}, Anne M. and {B{'o}di}, Attila and {Bapat}, Shreyas and {Barentsen}, Geert and {Baz{'a}n}, Juanjo and {Biswas}, Manish and {Boquien}, M{'e}d{'e}ric and {Burke}, D.~J. and {Cara}, Daria and {Cara}, Mihai and {Conroy}, Kyle E. and {Conseil}, Simon and {Craig}, Matthew W. and {Cross}, Robert M. and {Cruz}, Kelle L. and {D'Eugenio}, Francesco and {Dencheva}, Nadia and {Devillepoix}, Hadrien A.~R. and {Dietrich}, J{"o}rg P. and {Eigenbrot}, Arthur Davis and {Erben}, Thomas and {Ferreira}, Leonardo and {Foreman-Mackey}, Daniel and {Fox}, Ryan and {Freij}, Nabil and {Garg}, Suyog and {Geda}, Robel and {Glattly}, Lauren and {Gondhalekar}, Yash and {Gordon}, Karl D. and {Grant}, David and {Greenfield}, Perry and {Groener}, Austen M. and {Guest}, Steve and {Gurovich}, Sebastian and {Handberg}, Rasmus and {Hart}, Akeem and {Hatfield-Dodds}, Zac and {Homeier}, Derek and {Hosseinzadeh}, Griffin and {Jenness}, Tim and {Jones}, Craig K. and {Joseph}, Prajwel and {Kalmbach}, J. Bryce and {Karamehmetoglu}, Emir and {Ka{l}uszy{'n}ski}, Miko{l}aj and {Kelley}, Michael S.~P. and {Kern}, Nicholas and {Kerzendorf}, Wolfgang E. and {Koch}, Eric W. and {Kulumani}, Shankar and {Lee}, Antony and {Ly}, Chun and {Ma}, Zhiyuan and {MacBride}, Conor and {Maljaars}, Jakob M. and {Muna}, Demitri and {Murphy}, N.~A. and {Norman}, Henrik and {O'Steen}, Richard and {Oman}, Kyle A. and {Pacifici}, Camilla and {Pascual}, Sergio and {Pascual-Granado}, J. and {Patil}, Rohit R. and {Perren}, Gabriel I. and {Pickering}, Timothy E. and {Rastogi}, Tanuj and {Roulston}, Benjamin R. and {Ryan}, Daniel F. and {Rykoff}, Eli S. and {Sabater}, Jose and {Sakurikar}, Parikshit and {Salgado}, Jes{'u}s and {Sanghi}, Aniket and {Saunders}, Nicholas and {Savchenko}, Volodymyr and {Schwardt}, Ludwig and {Seifert-Eckert}, Michael and {Shih}, Albert Y. and {Jain}, Anany Shrey and {Shukla}, Gyanendra and {Sick}, Jonathan and {Simpson}, Chris and {Singanamalla}, Sudheesh and {Singer}, Leo P. and {Singhal}, Jaladh and {Sinha}, Manodeep and {Sip{H{o}}cz}, Brigitta M. and {Spitler}, Lee R. and {Stansby}, David and {Streicher}, Ole and {{{S}}umak}, Jani and {Swinbank}, John D. and {Taranu}, Dan S. and {Tewary}, Nikita and {Tremblay}, Grant R. and {Val-Borro}, Miguel de and {Van Kooten}, Samuel J. and {Vasovi{'c}}, Zlatan and {Verma}, Shresth and {de Miranda Cardoso}, Jos{'e} Vin{'i}cius and {Williams}, Peter K.~G. and {Wilson}, Tom J. and {Winkel}, Benjamin and {Wood-Vasey}, W.~M. and {Xue}, Rui and {Yoachim}, Peter and {Zhang}, Chen and {Zonca}, Andrea and {Astropy Project Contributors}},
                title = "{The Astropy Project: Sustaining and Growing a Community-oriented Open-source Project and the Latest Major Release (v5.0) of the Core Package}",
              journal = {\apj},
                 year = 2022,
                month = aug,
               volume = {935},
               number = {2},
                  eid = {167},
                pages = {167},
                  doi = {10.3847/1538-4357/ac7c74},
        archivePrefix = {arXiv},
               eprint = {2206.14220},
         primaryClass = {astro-ph.IM},
               adsurl = {https://ui.adsabs.harvard.edu/abs/2022ApJ...935..167A}
        }

@ARTICLE{Fuda2024,
       author = {{Fuda}, Nguyen and {Apai}, D{\'a}niel and {Nardiello}, Domenico and {Tan}, Xianyu and {Karalidi}, Theodora and {Bedin}, Luigi Rolly},
        title = "{Latitude-dependent Atmospheric Waves and Long-period Modulations in Luhman 16 B from the Longest Light Curve of an Extrasolar World}",
      journal = {\apj},
         year = 2024,
        month = apr,
       volume = {965},
       number = {2},
          eid = {182},
        pages = {182},
          doi = {10.3847/1538-4357/ad2c84},
archivePrefix = {arXiv},
       eprint = {2403.02260},
 primaryClass = {astro-ph.EP},
       adsurl = {https://ui.adsabs.harvard.edu/abs/2024ApJ...965..182F}
}

@ARTICLE{Miles_Paez2025,
       author = {{Miles-P{\'a}ez}, P.~A. and {Metchev}, S. and {Zapatero Osorio}, M.~R. and {Mart{\'\i}n-Carrero}, D.},
        title = "{Detection of J-band photometric periodicity in the T8 dwarfs 2MASS J09393548-2448279 and EQ J1959-3338}",
      journal = {\aap},
         year = 2025,
        month = may,
       volume = {697},
          eid = {L10},
        pages = {L10},
          doi = {10.1051/0004-6361/202555074},
archivePrefix = {arXiv},
       eprint = {2504.20672},
 primaryClass = {astro-ph.SR},
       adsurl = {https://ui.adsabs.harvard.edu/abs/2025A&A...697L..10M}
}

@ARTICLE{Zhang2021,
       author = {{Zhang}, Zhoujian and {Liu}, Michael C. and {Marley}, Mark S. and {Line}, Michael R. and {Best}, William M.~J.},
        title = "{Uniform Forward-modeling Analysis of Ultracool Dwarfs. I. Methodology and Benchmarking}",
      journal = {\apj},
         year = 2021,
        month = jul,
       volume = {916},
       number = {1},
          eid = {53},
        pages = {53},
          doi = {10.3847/1538-4357/abf8b2},
archivePrefix = {arXiv},
       eprint = {2011.12294},
 primaryClass = {astro-ph.SR},
       adsurl = {https://ui.adsabs.harvard.edu/abs/2021ApJ...916...53Z}
}

@ARTICLE{Marley_Ackerman,
   author = {{Marley}, M.~S. and {Ackerman}, A.~S.},
    title = "{The Role of Clouds in Brown Dwarf and Extrasolar Giant Planet Atmospheres}",
  journal = {ArXiv Astrophysics e-prints},
   eprint = {arXiv:astro-ph/0103269},
     year = 2001,
    month = mar,
   adsurl = {http://adsabs.harvard.edu/abs/2001astro.ph..3269M}
}
\bibliographystyle{aasjournal}



\end{document}